\documentstyle[prl,aps,epsfig]{revtex}
\draft
\begin{document}

\twocolumn[
\hsize\textwidth\columnwidth\hsize\csname@twocolumnfalse\endcsname

\title{Metal-insulator transition in the one-dimensional Kondo lattice model}
             
\author{Karyn Le Hur}
\address{Laboratoire de Physique des Solides, Universit{\'e} Paris--Sud,
		    B{\^a}t. 510, 91405 Orsay, France}
 \maketitle

\begin{abstract}
We study the usual one-dimensional Kondo lattice model (1D KLM) using the non-Abelian
bosonization. At half-filling, we obtain a Kondo insulator with a gap in
both charge and spin excitations which varies quite linearly with the Kondo
exchange $J_K$. It consists of a Spin Density Glass state, or a $q=\pi$ spin density wave
weakly pinned by a nearly antiferromagnetically ordered spin array. Away from
half-filling, the metallic system now yields a very small spin gap which
is equal to the one-impurity Kondo gap $T_k^{(imp)}$. Unlike the one-impurity 
Kondo model, we will show that this Kondo phase cannot rule the fixed point of 
the 1D KLM, away from half-filling. We rather obtain a normal heavy-fermion 
state controlled by the energy scale $T_{coh}\sim (T_k^{(imp)})^2/t$ (t is 
the hopping term) with a quite long-range antiferromagnetic polarization.

\end{abstract}

\vfill
 \twocolumn
\vskip.5pc ]
\narrowtext

\section{Introduction}

The Kondo lattice model (KLM) consists of conduction electrons, ruled
by the hopping term $t$, antiferromagnetically coupled to a spin array 
through the so-called Kondo interaction $J_K$\cite{un}. We are interested in this model for its possible applicability to the understanding of Kondo insulators like
$CeBi_3Pt_4$\cite{deux}, and heavy-fermion materials\cite{trois}. As a first 
step, the one-dimensional version of the KLM (1D KLM) has been studied by renormalization group methods\cite{quatre,cinq}, exact analytic methods\cite{six}, numerical 
diagonalization\cite{sept,huit}, density matrix renormalization group (DMRG)\cite{neuf}, non-linear sigma model\cite{neuf1} and bosonization methods\cite{dix,dix1,dix2,onze,onze1}. 

A large Kondo coupling forces all conduction electrons to form on-site singlets
with localized spins. Away from half-filling, the unpaired localized spins then
effectively hop around on the background of singlets and surprisingly interactions
in ${\cal O}(t^2/J_K)$ lead to incomplete ferromagnetism\cite{un,six}. On the other hand, the small coupling region $J_K/t$ is of particular interest since correlation
effects are most important and remains very hard to handle. When $J_K=0$, the conduction band is totally decoupled from the spin array to form a paramagnetic
band metal, resulting in gapless charge and spin excitations. However, the total
system is highly degenerate because of the completely free localized spins. Once switched on, the Kondo coupling should lift the spin degeneracy leading to
a complicated dynamics. Therefore, it is far from trivial whether the paramagnetic metallic phase at $J_K=0$ persists for finite $J_K$.

It has been rather clearly established that at half-filling, the
1D KLM, has a gap to both spin and charge excitations, leading to a 
Kondo insulator for any {\it non-zero} $J_K$\cite{quatre,six,neuf,d}. At na\" ive 
level, the tendency of these systems to form an insulator is a consequence of 
the Kondo effect, which admixes the spin scattering centers into the 
conduction sea, effectively counting each one as a quasiparticle. Kondo 
insulators are systems where, if one counts the spins of the array as 
electrons, there are an {\it even} number of electrons per unit cell. Those
systems are then characterized by a large Fermi surface. The Kondo process works as an effective on-site repulsion between conduction electrons. On the other side, such an insulator is
known to be ruled by a quite long-range antiferromagnetic ordering with
wave vector $q=\pi$\cite{neuf}. Interestingly, the staggered susceptibility $\chi(q=\pi)$
is predicted to diverge as $J_K\rightarrow J_{Kc}=0$, although $\chi(q=0)=0$ since the ground state is a singlet with a gap to spin excitation\cite{neuf}. Such a Kondo insulator does not belong to the well-known class of spin liquid states, where the spin correlation functions decay exponentially.

The behavior away from half-filling remains controversial. Numerical studies 
available until now, support the picture of a Tomonaga-Luttinger (TL) 
liquid\cite{douze1,douze2} with dominant correlations determined by the 
conduction electrons, which we may call ``RKKY liquid''
\cite{vingt,trente,douze}. But, at this moment, it should be mentioned
that system sizes used for the numerical studies are not sufficiently big
to determine the low temperature properties\cite{un}. On the other side, the Abelian
bosonization methods rather suggests, the possibility of an interesting
phase with a metallic behavior and a spin gap\cite{dix2}. Finally, since the essence of the so-called normal heavy-fermion state
lies in its quantum fluctuations, one-dimensional Kondo lattice
 should be a good test for understanding of the heavy-fermion state.

In this paper, we properly re-investigate the 1D KLM, using the non-Abelian bosonization methods\cite{onze,onze1}. Unlike the Abelian bosonization scheme\cite
{dix2}, we hope to perturb the model around the isotropic Kondo interaction 
limit. We mention that all the difficulty with this model is to include collective
effects occurring in the spin array at low-energy. Indeed, at high-energy, the
magnetic impurities behave as independent scattering centers. To help the reader, we recapitulate now the main steps and the important results found in this
paper.

\vskip 0.2cm
$\bullet$ At half-filling, we will obtain a precise
 Kondo insulator, namely the Spin Density Glass (SDG) state, with a gap both 
 in charge and spin sectors which varies linearly with $J_K$ for non-interacting
 electrons. For the charge gap, a $J_K$-linear behavior has been suggested in
 ref.\cite{zero} using DMRG methods. The resulting system consists of a $q=2k_F$ spin density wave (SDW) weakly 
 pinned by an antiferromagnetically ordered spin array. The appearance of 
 quite long-range order in the spin array shows evidence that the backward 
 Kondo scattering processes are very coherent, and finally leads to a Kondo 
 like localization of {\it all} the one-dimensional electron gas. A repulsive 
 interaction $U$ between electrons does not affect much the ground state. In 
 one dimension, quenched disorder is known to have a great influence on a 
 one-dimensional electron gas\cite{treize}, and in particular leads to the 
 so-called Anderson localization\cite{quatorze} in the non-interacting 
 case. We will analyze the stability of the SDG in presence of randomness. We 
 will compare these results with those already obtained in the 1D Heisenberg-Kondo lattice model (1D HKLM) where the spins of the array are now coupled through a large $J_H>>J_K$ Heisenberg Kondo 
exchange\cite{onze1}. 

\vskip 0.2cm
$\bullet$ Away from half-filling, we will prove the existence of a phase with a 
massless charge sector and a spin gap which is of the order in magnitude
of the one-impurity Kondo 
energy scale $T_k^{(imp)}\propto e^{-1/\rho J_K}$, where $\rho$ is the density 
of states per spin. But, unlike the one-impurity Kondo problem\cite{quinze,seize}, we will
show how such a phase cannot rule the fixed point of the Kondo lattice model
away from half-filling. We will rather discuss the necessity of two energy 
scales in the Kondo lattice away from half-filling, in view to engender a very 
low-energy coherence in the spin array and finally to obtain a heavy-fermion fixed point.

It is important to notice that both the SDG state and the heavy-fermion
state are susceptible to occur near a magnetic instability.

\section{Non-Abelian formulation of the model: a pedagogical step}

Our starting point is the Hamiltonian:
 \begin{eqnarray}
 \label{zero}
 H&=&-t\sum_{i,\sigma} c^{\dag}_{i,\sigma}c_{i+1,\sigma}+ (h.c) + U
 \sum_{i,\sigma}n^c_{i,\sigma}n^c_{i,-\sigma}\\ \nonumber
 &+&J_K\ c^{\dag}_{i,\alpha}({\sigma}^{\alpha}_{\beta}/2)c_{i,\beta}.\vec{\tau}_i
\end{eqnarray}
where the bare parameters obey $(U,J_K)<<t$. Here, $c^{\dag}_{i,\sigma}$ $(c_{i,\sigma})$ creates (annihilates) an electron of spin $\sigma$ at site i and
$\vec{\tau}_i$ is a spin $\frac{1}{2}$ operator located at site $i$. A small
U-Hubbard interaction between c-electrons has been introduced for more generality, and $J_K$ 
describes the usual Kondo coupling. 

In view to preserve the SU(2) spin 
symmetry, we use a continuum limit of the Hamiltonian and we switch over to 
non-Abelian bosonization notations\cite{onze,onze1}. The relativistic fermions $c_{\sigma}(x)$ are separated in 
left-movers $c_{L\sigma}(x)$ and right-movers $c_{R\sigma}(x)$ on the Fermi-cone. 
\begin{equation}
c_{i}\rightarrow \sqrt{a}c(x),\ c(x)=e^{-ik_F x}
 c_R(x)+e^{ik_F x}c_L(x)
 \end{equation}
 Since we are interested in low-energy properties, we may linearize the 
 dispersion of conduction electrons; the lattice step is now fixed to $a=1$. 
 The well-known {\it spin-charge separation} phenomenon takes place for 
 $U=0$, and the fermionic Lagrangian shows two different {\it conserved} currents for the 
 charge and spin degrees of freedom, namely $J_{c,L}=:c^{\dag}_{L\sigma}c_{L\sigma}(x):$ and 
 $\vec{J}_{c,L}=:c^{\dag}_{L\alpha}\frac{{\vec{\sigma}_{\alpha\beta}}}
 {2}c_{L\beta}:$ and similarly for the right-movers. As usual, we have ordered normally the current operators by subtracting any
  infinite constant. For non-interacting fermions $(U=0)$, the fermionic 
  Hamiltonian can be summed up in two separated Hamiltonians quadratic in 
  currents. 
 
 When $U<<t$, we expect the spin-charge separation phenomenon to
 survive. Using the conventional Abelian representation, the charge 
 Hamiltonian is equivalently described in terms of the
massless scalar field $\Phi_c$ and its moment conjugate $\Pi_c$:
\begin{eqnarray}
 (J_{c,L}+J_{c,R})&=&(1/\sqrt{\pi})\partial_x\Phi_c\\ \nonumber
 (J_{c,L}-J_{c,R})&=&(1/\sqrt{\pi})\Pi_c
 \end{eqnarray}
 It gives the precise Hamiltonian:
\begin{eqnarray}
 H_c&=&\int dx\  \frac{u_{\rho}}{2 K_{\rho}}:{(\partial_x\Phi_c)}^2:+\frac{u_{\rho}K_{\rho}}{2}: {(\Pi_c)}^2:\\ \nonumber
      &+&g_3\exp(i4k_Fx)\cos(\sqrt{8\pi}\Phi_c)
\end{eqnarray}
The coupling $g_3\propto U$ generates the usual $4k_F$-Umklapp process 
{\it only} relevant at half-filling, and the parameters $u_{\rho}$ and 
$K_{\rho}$ which describe the TL liquid are given by\cite{douze1,douze2}:
 \begin{equation}
  u_{\rho}K_{\rho}=v_F \qquad \text{and} \qquad \frac{u_{\rho}}{K_{\rho}}=v_F+2U/\pi
  \end{equation}
The Fermi velocity is $v_F=2t\sin k_F$. Maintaining the $SU(2)$ symmetry in 
the spin sector, it gives a k=1 Wess-Zumino-Witten (WZW) Hamiltonian\cite{a}:
\begin{equation}
 H_s=\frac{2\pi v_F}{3}\int\ dx\ :\vec{J}_{c,L}({x})\vec{J}_{c,L}({x}):+(L\rightarrow R)
\end{equation}
For $U=0$, the spin and charge degrees of freedom have the same velocity: physical
excitations are still particle-hole excitations. However, when $U\not=0$, the spin and charge
sectors behave independently one from another. Excitations are collective excitations, namely
{\it spinons} and {\it holons}. To treat the Kondo interaction, we need the complete representation
for the conduction spin operator\cite{onze}:
  \begin{eqnarray}
 \vec{S}_c&\simeq&\vec{J}_{c,L}(x)+\vec{J}_{c,R}(x)\\ \nonumber
 &+&\alpha\exp(i2k_F x)tr(g.\vec{\sigma})(x)\cos(\sqrt{2\pi}\Phi_c)(x)
 \end{eqnarray}
where $\alpha$ is a simple constant. Finally, the two relevant spin 
couplings come out as:
\begin{eqnarray}
&\lambda_2&(\vec{J}_{c,L}(x)+\vec{J}_{c,R}(x))\vec{\tau}(x)\\ \nonumber
&+&\lambda_3\exp(i2k_Fx)trg.\vec{\sigma}(x).\cos[\sqrt{2\pi}\Phi_c(x)]\vec{\tau}(x)
\end{eqnarray}
where $\lambda_{i=2,3}\propto J_K$ are dimensionless parameters. The SU(2) currents $\vec{J}_{c,L(R)}$ have scaling dimension 1, while the matrix field $g(x)$ and
the charge operator $\cos[\sqrt{2\pi}\Phi_c(x)]$ have respectively the scaling
dimensions 1/2 and $K_{\rho}/2$. The distance between two magnetic impurities
is $a=1$. Then, we may also take the continuum limit with respect to the impurity lattice. The {\it discrete} character of the spin array can be ignored, and
$\vec{\tau}_i\rightarrow a\vec{\tau}(x)$\cite{dix2}. It will be useful to 
describe the coherence between spins of the array. We insist on the
fact that it is very difficult to consider theoretically the possible low-energy ``dynamics'' in the spin array, because magnetic impurities
are completely decoupled at high energy.

\section{Commensurate filling}

We investigate the recursion equations of $\lambda_2$ and
$\lambda_3$ up to the order $J_K^2$. The current operators obey the so-called
$SU(2)_{k=1}$ Kac-Moody algebra\cite{onze}:
\begin{eqnarray}
[J_{c,L}^a(z),J_{c,L}^b(z')]&=&if^{abc}J_{c,L}^c(z)\delta(z-z')\\ \nonumber
&+&\frac{i}{4\pi}\delta^{ab}\delta'(z-z')
\end{eqnarray}
with $f^{xyz}=1$, and $z=(x+v_Ft)$ is the Fermi-cone 
component. The impurity spin operators behave as {\it hard-core} boson ones
and follow the Lie algebra:
\begin{equation}
[{\tau}^a(x),{\tau}^b(x')]=if^{abc}{\tau}^c(x')\delta(x-x')
\end{equation}
They \underline{commute} on different sites. Consequently, using the equation:
\begin{equation}
\frac{1}{v_F}\int \frac{du}{u}\sim \frac{ln L}{v_F}\qquad u=v_F.t
\end{equation}
we find, that $\lambda_2$ obeys the following beta function:
\begin{equation}
\beta(\lambda_2)=\frac{d\lambda_2}{dln L}=\frac{\lambda_2^2}{2\pi v_F}
\end{equation}
We realize, that it is completely equivalent to the beta function found in 
the usual one-impurity Kondo model\cite{quinze,seize}. The exchange coupling $\lambda_2$, which involves the $q=0$ spin density operator, scales to the strong coupling regime at the {\it one-impurity} Kondo energy scale $T_k^{(imp)}\propto 
E_c e^{-2\pi v_F/J_K}$, $E_c\sim t$ is the bare bandwidth cut-off. In that sense, $\lambda_2=\lambda_2 \delta(x)$ and no correlation is expected
in the spin array up to the energy scale $T\sim T_k^{(imp)}$. 

But, at half-filling, the $q=2k_F$ spin density wave becomes commensurate with 
the lattice, and the term $\lambda_3$ which generates the backward 
Kondo scattering term should be strongly more relevant than the forward Kondo 
scattering term. To prove that explicitly, we adopt the following scheme. Due to the ${\bf (-1)^x}$ factor in the 
term $\lambda_3$, the ground state is expected to be one in which the impurity spins $\vec{\tau}(x)$ are 
antiferromagnetically coupled. They would obey $\langle 
\vec{\tau}(x)\vec{\tau}(x')\rangle\sim const.$, and should exhibit a (quite)
long-range order at $T\rightarrow 0$. This is the non-local order parameter 
which will define at half-filling the following coherent state: the Spin
Density Glass (SDG) state. Indisputably, in 1D, no true
long-range order can persist, but quantum spin fluctuations would be minimized 
at the fixed point.

\subsection{Spin density glass (SDG) and magnetic instability}

The staggered operator $\lambda_3$ has a zero conformal spin. Then, to
obtain the recursion law of $\lambda_3$, it is enough to treat 
$\{\tau^i(x)\}$ (i=x,y,z) as c-numbers ($\tau^i(x)=(-1)^x\frac{1}{2a}$), labeling eigenstates of the 
Hamiltonian with {\it static} configuration of $\tau^i(x)$. The chosen representation
indicates that no own excitation occur in the spin array, and that an antiferromagnetic
polarization is created by the staggered operator $\lambda_3$. Then, to make the partition function ${\cal Z}$ invariant under the cut-off 
transformation $a=1\rightarrow a'=e^{dln L}$, the term $\lambda_3$ has to 
obey: $\lambda_3^2(a')=\lambda_3^2(a=1)(\frac{1}{a'})^{(3-K_{\rho})}$. It gives the important equation:
\begin{equation}
\label{a}
\beta(\lambda_3)=\frac{d\lambda_3}{dln L}=\frac{1}{2}(3-K_{\rho})\lambda_3
\end{equation}
We obtain, that the backward Kondo scattering process 
$\lambda_3$ becomes strong at the energy scale:
 \begin{equation}
 T_k^{(3)}\propto E_c(J_K/t)^{2/(3-K_{\rho})}>>T_k^{(imp)}
 \end{equation}
Indisputably, it dominates the low-energy physics, for a commensurate 
filling. Since it couples spin and charge degrees of freedom, this produces 
both a charge gap and a spin gap of the order of $T_k^{(3)}$. In fact, this conclusion is independent on the sign of $J_K$. Indeed, equation (\ref{a}) remains unchanged by the exchange $\lambda_3\rightarrow -\lambda_3$. However, the 
used formalism does not allow to conclude that the charge gap is larger than 
the spin one\cite{quatre,d}. The well-known spin-charge separation phenomenon disappears for any finite value of $J_K$. In the case of non-interacting 
electrons, $T_k^{(3)}$ varies linearly with $J_K$ (cf Fig.1), as in the strong
Kondo coupling regime. The Kondo effect plays the role of a strong on-site 
repulsive interaction between electrons. We also remark, that it increases with $U$ (cf Fig.1); the electrons become more localized and the ground state still more stable. $4k_F$-Umklapps become {\bf useless} in the model; they would affect the physics at an energy scale $\sim E_c e^{-\pi v_F/U}<<T_k^{(3)}$ for $(U,J_K)<<t$. Finally, due to the cosine term in $\Phi_c$, the renormalized exponent 
$K_{\rho}^*$ tends to 1/2 when $T<<T_k^{(3)}$. But, $K_{\sigma}^*=1$ since
the Kondo exchange is supposed to be invariant per rotation. To understand 
more physically the ground state and excitation spectra, we use 
the Abelian representation of $trg\vec{\sigma}(x)$:
\begin{equation}
[\sin(\sqrt{2\pi}\tilde{\Phi}_s),-\cos(\sqrt{2\pi}\tilde{\Phi}_s),\sin(\sqrt{2\pi}{\Phi}_s)]
\end{equation}
where ${\Phi}_s$ is the spin field and $\tilde{\Phi}_s$ its dual field. For
a static configuration of the spin array, the
term $\lambda_3$ can be re-expressed as the two Sine-Gordon terms:
\begin{eqnarray}
& &\int dx\ \frac{\lambda_{3z}}{a^2}\sin[\sqrt{2\pi}{\Phi}_s(x)]\cos[\sqrt{2\pi}{\Phi}_c(x)]\\ \nonumber
&+&\frac{\lambda_{3\perp}}{a^2}\exp[-i\sqrt{2\pi}\tilde{\Phi}_s(x)]\cos[\sqrt{2\pi}{\Phi}_c(x)]+(h.c)
\end{eqnarray}

\vskip 0.2cm
$\bullet$ {\it Coherent ground state}: The Hamiltonian is minimized when the fields $\Phi_c$ and $\Phi_s$ respectively take the values $(2n+1)\sqrt{\pi/2}$ and $(2n+3/2)\sqrt{\pi/2}$, and $n$ is an integer. The ground state resembles a $q=\pi$ spin density wave weakly pinned by an antiferromagnetically ordered spin array; we confirm a 
pioneer class of one-dimensional Kondo insulators, namely the SDG 
(``Spin Density Glass'') state. The backward Kondo scattering processes are very {\it coherent}, and leads to a 
magnetic like localization of the one-dimensional electron gas. Therefore, the
conduction electrons are subject to a quasi-static potential, as if there is
a staggered spin moment. Finally,
the Kondo coupling allows to lift the high degeneracy of the impurity
lattice by minimizing the dynamics in the spin array. 

In the case of non-interacting electrons, density-density correlation 
functions decay as $1/x^2$. As could expected, the single enhanced order 
parameter (defining the electron gas) which does not decay 
exponentially is the $\pi$-SDW; $O_{SDW}(x)=[c^{\dag}_{L\uparrow}
c_{R\downarrow}+(h.c)](x)$. At long distances: $\langle O_{SDW}(x)O_{SDW}(0)\rangle\sim \frac{(-1)^x}{x}$ as for a Heisenberg chain. The one-dimensional electron gas has inevitably enhanced $q=\pi$ spin correlations. The $\pi$-charge density wave $O_{CDW}(x)=[c^{\dag}_{L\uparrow}c_{R\uparrow}+c^{\dag}_{L\downarrow}c_{R\downarrow}](x)\propto\cos[\sqrt{2\pi}\Phi_s(x)]\cos[\sqrt{2\pi}\Phi_c(x)]\rightarrow 0$ because the ground state obeys $\Phi_s=(2n+3/2)\sqrt{\pi/2}$. This last result does not seem in agreement with the authors of ref.\cite{dix2}. In fact, they have used a particular solvable point $\lambda_{2z}=v_F\pi$, at which the backward Kondo scattering term has no real influence. But, since the term $\lambda_3$ flows to the strong
coupling regime before $\lambda_2$, we conclude that the used solvable point
has not a real significant meaning at half-filling.

\vskip 0.2cm
$\bullet$ {\it Excitation spectra}: The quantum numbers of the excitations are 
easily determined with the Abelian bosonization formalism. Solitons in the 
$\Phi_s$ field are $\sqrt{2\pi}$ phase slips, amounting to a state with spin 
S=1 and charge $Q=1$ (Fig.2a). It implies the deconfinement of spinons in
the TL liquid. Similarly, a $\Phi_c$ soliton carries $S=0$ and $Q=2$.

\subsection{Short localization length}

The coherent state vanishes for a critical length $\xi_{c}\sim v_F/T_k^{(3)}$. For a small $J_K$ coupling, Eq.(\ref{a}) gives:
\begin{equation}
\xi_{c}\sim [\frac{1}{J_K}]^{\frac{2}{3-K_{\rho}}}
\end{equation}
It is the (small) length at which we must stop renormalization. Since $\xi_c$
is not large, there is no consequent overlap between spin singlets on neighboring sites at the fixed point $(L\rightarrow +\infty)$. Now, we are rather
interested in the delocalized phase for $L<\xi_{c}$. A small term $\lambda_3$ changes the direction of propagation of particles, and it would modify the universal conductivity of the electron 
gas\cite{douze2}:
\begin{equation}
\sigma=2e^2K_{\rho}L/h 
\end{equation}
obtained by applying a static field over a finite part of the sample without
magnetic impurities. The current DC conductivity obeys the scale invariance
law:
\begin{equation}
\sigma(T)=\sigma_o T^{-1}\frac{D_{mo}}{D_m(T)}
\end{equation}
where $D_m\sim\lambda_3^{2}$ generates the magnetic disorder, and the sign $o$
refers to the fixed bare conditions. It gives:
\begin{equation}
\sigma(T)\propto T^{-1}\lambda_3(T)^{-2}\sim T^{2-K_{\rho}}
\end{equation}
Due to the influence of the staggered operator $\lambda_3$, the spin
array is attempted to yield a very short-range antiferromagnetic 
polarization. Then, we can check that (Appendix A):
\begin{equation}
\sigma(T)>\delta G_{imp}^{-1}(T)
\end{equation}
where $\delta G_{imp}\propto T^{K_{\rho}-1}$ is the variation of conductance 
of the one-dimensional electron gas in presence of a single magnetic 
impurity. It shows that short range spin correlations 
in the array rather enhance the DC conductivity, compared to the case where the spins were totally independent. Scatterings with impurities become 
important only when $T\rightarrow T_k^{(3)}$, leading to a sharp cross-over 
towards a localization phase: a long-range antiferromagnetic polarization in the spin array takes place. It proves the important 
result, that the SDG state can only occur near a magnetic instability.

Finally, we suggest that the SDG remains quite stable by deleting a single impurity
from the periodic array. More precisely, the introduction of a single Kondo hole
is expected to generate a localized spin S=1/2, which should bring a Curie
component in the magnetic spin susceptibility\cite{d}. This conclusion is due to the
fact that the free spin S=1/2 introduced in the conduction band feels a charge
gap of order $T_k^{(3)}$ imposed by the structure, but no spin gap. The Kondo
effect plays the role of a strong on-site electronic repulsion.

\subsection{The Heisenberg-Kondo lattice: enhancement of quantum fluctuations}

Now, we compare this result with this obtained in the so-called 
Heisenberg-Kondo chain\cite{onze1}, in which the impurity spins are coupled through a large Heisenberg exchange term $J_K<<(J_H,t)$. In that case, it is convenient to use the following
representation:
\begin{equation}
\vec{\tau}(x)\simeq\vec{J}_{f,L}(x)+\vec{J}_{f,R}(x)+\text{constant}.(-1)^x tr(f.\vec{\sigma})
 \end{equation}
It has been well established, that the low-energy physics is still ruled by
the term $\lambda_3$; but here, it obeys the following recursion equation:
\begin{equation}
\beta(\lambda_3)=\frac{d\lambda_3}{dln L}=\frac{1}{2}(2-K_{\rho})\lambda_3
\end{equation}
It still opens a mass gap $T_k^{(3)}\propto E_c(J_K/t)^{2/(2-K_{\rho})}$ for
the charge and all the spin excitations.
\\
\begin{figure}[ht]
\centerline{\epsfig{file=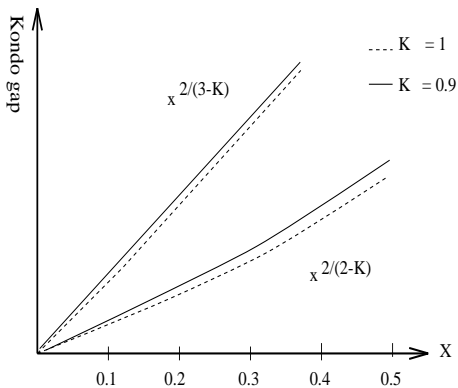,angle=0.0,height=5cm,width=6cm}}
\vskip 1cm
  \caption{Kondo gap for $J_H=0$ and $J_H=t$, and for two different values
of the Hubbard interaction. We have used the notations $x=J_K/t$ and 
$K=K_{\rho}$.}
 \end{figure}
 
However, the ground state is quite
different from that obtained above for $J_H=0$ and in particular yields a
smaller gap in both charge and spin excitations (cf Fig. 1). It rather resembles the disordered quantum spin liquid state of the two-leg spin 
ladder system\cite{onze1}. It is known that low-energy excitations consist of a triplet branch
S=1, and at higher energy of a singlet continuum. In Appendix B, we propose an effective action describing well such an insulator, for $J_H\sim t$:
\\
\begin{equation}
\label{un}
S_{eff}=\frac{2}{4 u^2}\int d {\bf x}\ tr\partial_{\mu}g \partial^{\mu}
g^{\dag}-\gamma{[T_k^{(3)}]}^{3/2}trg
\end{equation}
\\
$\gamma=1/2$, $u^2\propto v_F^{-1}$ and ${\bf x}=(x,v_Ft)$. Here, the 
u-coupling becomes asymptotically free:
 \begin{equation}
 \beta(u)=\frac{\partial u^2}{\partial lnL}=\frac{1}{4\pi}u^4
 \end{equation}
It means that the ground state is confined inside the characteristic length:
   \begin{equation}
   \xi_H\propto \frac{u^2}{T_k^{(3)}}\exp(\frac{+4\pi}{u^2})
   \end{equation}
It confirms the loose of the long-range magnetic order due to incoherent
quantum fluctuations in the spin array. Here, the spin array has its own 
dynamics, and the deconfinement of spinons takes place inside each spin 
system. The ground state is non-degenerate; it is immediate to observe that 
site Parity $x\rightarrow -x$ is not broken at an energy $T<<T_k^{(3)}$ 
($trg=0$). We remind that site Parity acts on g as $g\rightarrow -g^{\dag}$. 

\vskip 0.2cm
$\bullet$ {\it Ground state and enhanced order parameters}: We finally obtain the same short-range Resonating Valence Bond (RVB) state which does not possess any free spinon on a length $\xi_H$, as in the purely two-leg spin ladder system. However, the spin gap increases considerably by tuning $K_{\rho}$ from 1 (for a free electron gas) to
0 (for a Heisenberg chain). When $K_{\rho}\rightarrow 0$, we recognize a
spin gap linear in the exchange coupling $\lambda_3$, that is typical of
the two-leg spin ladder system. The charge degrees
of freedom are completely frozen, that makes the RVB ground state more
stable.
\\
Unlike the case $J_H=0$, we can check that the staggered operator $(-1)^x trg\vec{\sigma}$ cannot be relevant at the fixed point due to the presence of the
term $trg$ in the effective action. The only order parameter which does not
decay exponentially is the $2k_F$ CDW order: 
$O_{CDW}=[c^{\dag}_{L\uparrow}c_{R\uparrow}+c^{\dag}_{L\downarrow}
c_{R\downarrow}]=trg(x)\cos[\sqrt{2\pi}\Phi_c(x)]$ 
(in the non-Abelian bosonization formalism). At long 
distances, $\langle O_{CDW}(x)O_{CDW}(0)\rangle\sim const.$, since the field 
$\Phi_c$ is gaped and the ground state yields $trg\not=0$. In the spin 
sector, correlations are inevitably provided by massive spin S=1 
objects. To find their precise correlation functions, we have to use the 
following {\it traceless} representation:
\begin{equation}
g=i\vec{\sigma}.\vec{n}\qquad  \vec{n}^2=1
\end{equation}
Then, the effective action (\ref{un}) becomes equivalent to the so-called 
O(3) non-linear $\sigma$-model\cite{b}. Excitations of this model are indisputably S=1 particles. The correlation functions are known to come out as: $\langle
\vec{n}(x,t)\vec{n}(0,0)\rangle\sim K_o(T_k^{(3)}r)+{\cal O}[exp(-3T_k^{(3)}r)]$ and $K_o$ is a Bessel function\cite{b}. Finally, $T_k^{(3)}$ can be identified with the so-called {\it Haldane} gap.
\\
\begin{figure}[ht]
\centerline{\epsfig{file=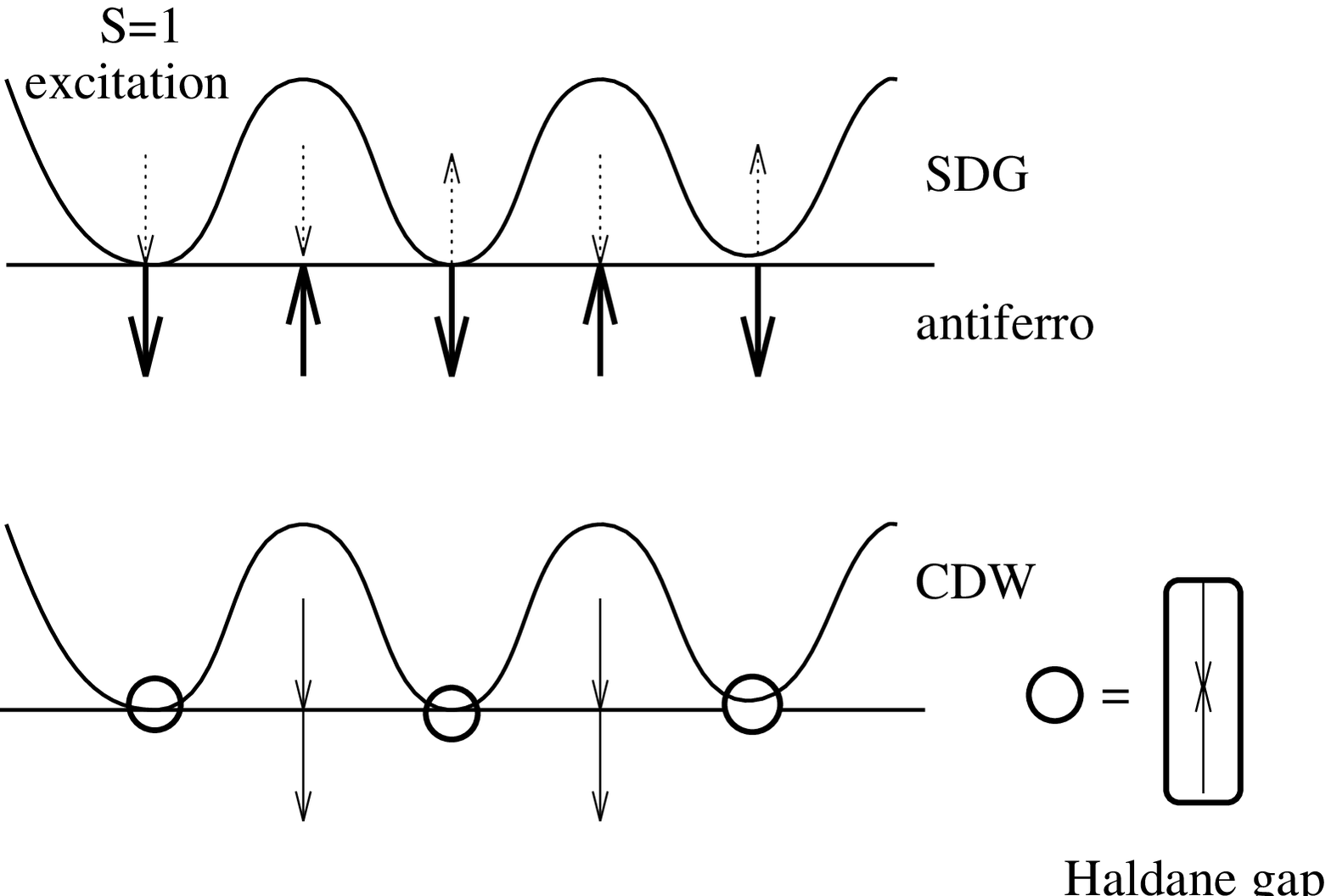,angle=0.0,height=4.5cm,width=8.5cm}}
\caption{Schematic pictures of the SDG state (a) and the Heisenberg-Kondo
 ground state (b). The range of the magnetic order in the spin array
 considerably decreases by increasing the Heisenberg exchange $J_H$.}
 \end{figure}

In the weak-coupling regime, we can check that the Heisenberg chain gives a
DC conductivity $\sigma(T)\propto T^{-1}\lambda_3^{-2}(T)\sim 
T^{1-K_{\rho}}$. The spin array develops its own excitations (namely spinons)
for $T<<J_H$, and finally the impurities act independently one from the other
in the local scattering processes. The small Kondo effect rather disturbs
the confinement of spinons in the Heisenberg chain.

\subsection{Influence of quenched disorder}

We have shown evidence of the effects of coherent scattering from many 
magnetic impurities which typically give rise to the ``Kondo
localization'' (for $J_H=0$). Now, we address the question: is this
coherent regime survive to the presence of many non-magnetic impurities? 
\\
\begin{figure}[ht]
\centerline{\epsfig{file=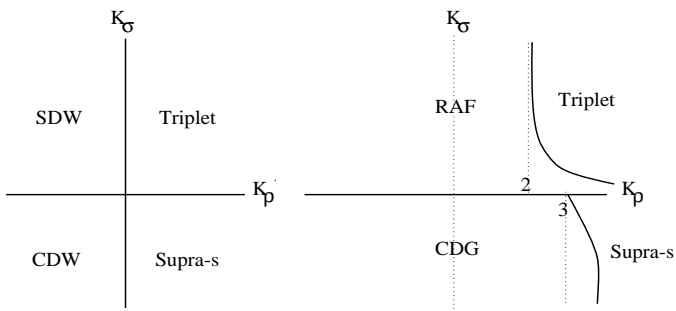,angle=0.0,height=4cm,width=9cm}}
\vskip 1cm
  \caption{(a) Phase diagram of the Tomonaga-Luttinger liquid. (b) Influence
of quenched disorder.}
 \end{figure}

Indeed, it is well-known that a one-dimensional electron gas is strongly submitted
to the influence of disorder; the effects of coherent scattering from many
non-magnetic impurities lead to the well-known ``Anderson localization'' in
the non-interacting case\cite{treize,quatorze}. We find it useful to remind the phase diagram of the TL liquid, and the influence of disorder on it 
(cf Figs.3)\cite{douze2} (last ref.). Finally, we find it very interesting to 
reconsider the phase diagram of the TL liquid in presence of both magnetic 
impurities and quenched disorder. To discuss the role of random impurities on the SDG state, it is convenient to 
use the conventional Abelian bosonization scheme. For the standard renormalization group methods, we refer the reader to the article of Giamarchi and 
Schulz\cite{treize} (which discusses disorder in a TL liquid).

We introduce the \underline{complex} random impurity potential,
 \begin{equation}
   H_{imp}=\sum_{\sigma}\int dx\ \xi(x)c^{\dag}_{\sigma L}c_{\sigma R}\qquad 
+(h.c)
   \end{equation}
   with the Gaussian distribution of width $D_{\xi}=n_i\xi(q=2k_F)^2$:
   \begin{equation}
   P_{\xi}=\exp(-D_{\xi}^{-1}\int dx\ {\xi}^{\dag}(x){\xi}(x))
   \end{equation}
$n_i$ is the small non-magnetic impurity density. We can omit forward 
scatterings, because the q=0 random potential just renormalizes the chemical 
potential and it does not affect the $2k_F$
SDW. $4k_F$-Umklapps can be forgotten too since they do not have any
influence on the charge sector, in presence of the backward Kondo scattering
term $\lambda_3$. 

In its bosonic form, the Hamiltonian $H_{imp}$ comes out as:
  \begin{equation}
   H_{imp}\simeq\frac{1}{2\pi a}\int dx\ \xi(x)\exp^{i(\sqrt{2\pi}\Phi_c-2k_Fx)}\cos\sqrt{2\pi}\Phi_s(x)
   \end{equation}
To treat commonly the quenched disorder, we use the famous {\it replica} 
trick. But, due to the restrictive definition of $\xi(x)$, we are limited to
a first order contribution treatment in $D_{\xi}$. There is no coupling between different replica indices, which will be omitted 
below. If we want to treat precisely the cases where the
interaction between electrons is rather attractive, we have to take also into account 
the usual Sine-Gordon term:
\begin{equation}
H_{ss}=\int dx\ g_1\cos[\sqrt{8\pi}\Phi_s(x)]
\end{equation}
with $g_1\propto U/(2\pi a^2)$. Finally, up to lowest orders in $D_{\xi}$, $\lambda_3$ and $g_1$, one now obtains a set of coupled renormalization group equations:
 \begin{eqnarray}
   \label{dad}
    \frac{d D}{dl}&=&(3-K_{\sigma}-K_{\rho})D-yD\\ 
    \label{deuxeee}
    \frac{dy_{3z}}{dl}&=&(2-\frac{K_{\sigma}}{2}-\frac{K_{\rho}}{2})y_{3z}\\
    \label{deuxe} 
    \frac{dy_{3\perp}}{dl}&=&(2-\frac{1}{2K_{\sigma}}-\frac{K_{\rho}}{2})y_{3\perp}\\ 
 \frac{dy}{dl}&=&2(1-K_{\sigma})y-D\\    
\label{deuxee}
    \frac{d K_{\sigma}}{dl}&=&-\frac{1}{2}(D +y_{3z}^2-y_{3\perp}^2+y^2)K_{\sigma}^2\\
   \label{deux}
    \frac{d K_{\rho}}{dl}&=&-\frac{u_{\rho}}{2u_{\sigma}}[D+y_{3z}^2+y_{3\perp}^2]K_{\rho}^2
   \end{eqnarray}
   with the notations: $dl=d\text{Ln} L$, $D=2\frac{D_{\xi}
   a}{\pi u_{\sigma}^2}[\frac{u_{\sigma}}{u_{\rho}}]^{K_{\rho}}$
   , ${y}_{3\nu}=\frac{\lambda_{3\nu}}{2\pi u_{\sigma}}$ and 
${y}=\frac{g_1}{2\pi u_{\sigma}}$, and where we
   did not display equations irrelevant to the following discussions. These
equations are valid for arbitrary $K_{\nu}$. We can remark that no term like
$Dy_{3z}$ is generated perturbatively; no anisotropy between 
$y_{3z}$ and $y_{3\perp}$ is engendered by randomness.

As a first application of such a flow of renormalization, one can determine
the phase diagram represented in Fig. 4a. One can already observe that the
point $K_{\rho}=1$ and $K_{\sigma}=1$ is {\it unstable}. In presence of
both magnetic and non-magnetic impurities, the free electron gas cannot
yield a stable fixed point. More generally, there are four different phases. 
\\
\begin{figure}[ht]
\centerline{\epsfig{file=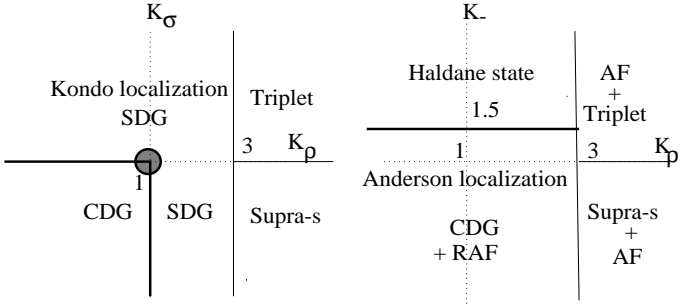,angle=0.0,height=4cm,width=9cm}}
\vskip 1cm
 \caption{Phase diagram of the Tomonaga-Luttinger liquid in presence of
both magnetic impurities and randomness. (a) Situation for $J_H=0$ established
with the above renormalization flow. (b) Situation for $J_H>>J_K$ obtained
in last ref.[15].}
 \end{figure}

1. For $K_{\sigma}<1$ and $K_{\rho}<1$: The pure TL liquid shows rather an 
enhanced $2k_F$ CDW (cf Fig.3a). We hope that the backward scattering processes
with the non-magnetic impurities become crucial at low-energy. In that case, we may
check that the disorder estimator $D$ flows to strong couplings before
the term $\lambda_{3z}$: we expect the so-called Anderson 
localization. The localization length due to randomness becomes very small:
\begin{equation}
\xi_r\sim [\frac{1}{D}]^{\frac{1}{3-K_{\sigma}-K_{\rho}}}<\xi_c
\end{equation}
Although a precise strong coupling treatment is not possible, we deduce that 
the fixed point is then ruled by $D^*\rightarrow +\infty$, $y^*\rightarrow 
-\infty$. This localized region can therefore be identified as a weakly pinned 
CDW, also called the ``Charge Density Glass'' (CDG) (cf Fig.3b). Since 
$y^*\rightarrow -\infty$, the CDG is a nonmagnetic spin singlet, where 
localized single-particle states are doubly occupied. This description is 
available only when the repulsion between electrons is not too strong.
\\

2. For $K_{\sigma}<1$ and $K_{\rho}>3$: One can remark that both
$D^*\rightarrow 0$ and $\lambda_{3\nu}^*\rightarrow 0$. This is a delocalized region ruled by $y^*\rightarrow -\infty$. There is a spin gap due to predominant
Singlet Superconductor (SS) fluctuations.
\\

3. For $K_{\sigma}>1$ and $K_{\rho}>3$: all $D$, $y$ and $y_3$ tend to zero
at the fixed point. This delocalized state favors triplet superconductivity.   
\\

4. In all other cases, backward scatterings with the impurity lattice 
remain prevalent. Using Eqs. (\ref{dad}), (\ref{deuxeee}) and (\ref{deuxe}), we check that the term
$\lambda_{3\perp}$ and $\lambda_{3z}$ scale to the strong coupling regime 
before the disorder parameter $D$. The Kondo localization still takes 
place because quantum fluctuations are considerably minimized in the spin 
array and $\xi_c$ is small. Then, although the weak-coupling analysis breaks 
down in this limit, we may predict that the fixed point should be governed by 
$\lambda_{3\nu}^*\rightarrow +\infty$, $K_{\sigma}^*=1$ and $D^*\rightarrow 0$.
\\

We can compare this phase diagram with that obtained by introducing randomness
in the 1D Heisenberg-Kondo lattice (cf Fig.4b). The disordered spin liquid 
state obtained at half-filling rather favors a $2k_F$ CDW. In presence of randomness, it must fight against the occurrence of the so-called CDG state. 
Anyway, the pinning of a CDW by a disordered quantum spin state is supposed
to be less stable than the Anderson localization by disorder. Charge
excitation gap is quite small. Then, it may explain why the Anderson 
localization remains prominent in a large region of the phase 
diagram $(K_-,K_{\rho})$. $K_-$ is the spin parameter which controls the 
high-energy singlet mode $\rho$ (cf Appendix B). Shifting the singlet branch to infinity makes the spin liquid state more ordered: quantum fluctuations are minimized, and 
the Anderson localization can be finally suppressed. Spin excitations are now
reduced to the classical ones $S^z=\pm 1$. The problem becomes completely equivalent
to that of two coupled Ising chains, and the topological order parameter which
characterizes the disordered spin system remains finite\cite{onze1,b}:
\begin{equation}
lim_{\left|i-j\right|\rightarrow +\infty}\langle S_i^z\exp(i\pi\sum_{k=i-1}^{j-1}
S_k^z)S_j^z\rangle=\langle\cos\sqrt{\pi}\Phi_+\rangle
\end{equation}
$S$ are spin-1 objects. In the Anderson localization phase, there is an entire separation between the one-dimensional gas pinned by randomness, and the Heisenberg chain which 
resembles more an antiferromagnet with random exchange (RAF).

\section{Incommensurate limit}

Even for a very small hole-doping, the $2k_F$ oscillation is not commensurate
with the lattice, and the term $\lambda_3$ can be dropped out\cite{onze1}. A new kind of metal-insulating transition is expected to take place.

At half-filling, we
remind that the charge sector can be modeled by the following Hamiltonian:
\begin{eqnarray}
 H_c&=&\int dx\  \frac{u_{\rho}}{2 K_{\rho}}:{(\partial_x\Phi_c)}^2:+\frac{u_{\rho}K_{\rho}}{2}: {(\Pi_c)}^2:\\ \nonumber
      &+&\nu\ e^{2ik_Fx}(-1)^x[T_k^{(3)}]^{3/2}\cos(\sqrt{2\pi}\Phi_c)
\end{eqnarray}
$\nu$ is a simple constant. If we define incommensurability $q$ such as $2(k_F
+q)=\pi$, the charge sector is expected to become massless for $q>q_c\sim
\pi/(2L_c)$, where 
\begin{equation}
L_c\sim \frac{a}{2\pi}\sqrt{\frac{\pi v_F}{\nu[T_k^{(3)}]^{3/2}}}
\end{equation}
is the soliton length of the Sine-Gordon term. In the incommensurate limit, the forward Kondo scattering term becomes then 
relevant and the perturbation theory fails under the usual one-impurity Kondo 
energy scale: $T_k^{(imp)}\propto E_c e^{-2\pi v_F/J_K}<<T_k^{(3)}$. In the following, the 
c-electrons are supposed to be {\it non-interacting}, and 
$K_{\rho}\rightarrow 1$. We treat the case where excitations in the electron
gas are particle-hole excitations.

Now, since the system is metallic, only the fermions 
very close to the Fermi level are expected to participate in the $q=0$ Kondo 
process. They obey:
\begin{equation}
\epsilon(k)=v_F\left|k-k_F\right|<<T_k^{(imp)}
\end{equation}
Indeed, the renormalization group approach integrates out the high energy 
degrees of freedom. Away from half-filling, a small number of conduction 
electrons participate in the Kondo process. Per site, we just dispose of:
\begin{equation}
n_c^k=\frac{T_k^{(imp)}}{t}<<1
\end{equation}
conduction electron to screen a magnetic impurity. We observe the 
well-known ``exhaustion'' phenomenon\cite{c}. 

Therefore, one can suggest two possible antagonist senarii concerning
the possible polarization of the spin array by conduction electrons, and
more generally the fixed point properties.
\\

1- One can choose to forget the {\it rare} Kondo singlets to analyze the spin
correlation properties: the low-energy physics is supposed to be governed by 
{\it unpaired} localized spins and c-electrons. In that case, one obtains a free electron gas defined by the density
$n_{eff}=(1-n_c^k)$ and the Fermi velocity $v_F=2t\sin\frac{\pi}{2}n_{eff}$, weakly coupled
to the non-screened spins of the array. The $q=0$ part of the
spin sector is redefined as:
\begin{equation}
\vec{J}_L\rightarrow \vec{J}_{c,L}+\frac{3\lambda_2^o}{2\pi v_F}\vec{\tau}
\end{equation}
$\lambda_2^o$ refers to the bare Kondo exchange. Since $\lambda_2^o<<v_F$ a
very small part of the non-screened spins delocalize. Finally, ``high-energy'' 
electrons generate prevalently the so-called $q=2k_F$ RKKY interaction ruled by$(a\lambda_2^{o})^2/t=J_K^2/t$. This result seems quite be well supported by 
numerical calculations\cite{un,vingt,trente}.
\\

2- In this paper, we rather argue that it is the Kondo process which generates
the {\it strongest} correlations in the spin array. In fact, as in the usual 
one-impurity Kondo model, the characteristic energy scale:
\begin{equation}
\xi_i=\frac{v_F}{T_k^{(imp)}}<<L=\frac{1}{k-k_F}
\end{equation}
must be taken into great account in the low-energy description. It is {\it considerably} larger than $\xi_c$ obtained at 
half-filling. Quantum fluctuations imposed by the Kondo screening process
should be the most prominent. Finally, all the low-energy 
properties should be induced by 
electrons near the Fermi level. To show that, we will omit the high-energy
electrons in the following.

First, in Appendix C, we find it useful to remind why a heavy-fermion fixed point occurs in the one-impurity Kondo problem (also submitted to a metallic 
behavior and the spin gap $T_k^{(imp)}$). Second, $\xi_i$ also occurs as a 
relevant energy scale in the Kondo lattice model away from half-filling. It is then necessary to 
establish its exact physical meaning, but also to
analyze possible {\it irrelevant} operators (cf Appendix C) in this 
fascinating lattice model. 

\subsection{Localization cloud in the Kondo lattice away from half-filling}

It is immediate to observe that $\xi_i$ has not the same meaning 
as in the one-impurity Kondo model. From another point of view, we rather observe 
that $z=1/n_c^k$ magnetic impurities tend to interact with the same conduction 
electron. The notion of screening cloud around a magnetic impurity has to be 
abolished. We also remark, that the situation is completely different from 
that described at half-filling, where the system was insulating and finally 
one electron per site $n_c^k=1$ was free to interact with a spin of 
the array. No irrelevant operator was susceptible to occur in that 
context. Here, it is important to notice that:
\begin{equation}
\xi_i=za=\frac{a}{n_c^k}
\end{equation}
Then, since the characteristic time ${\tau_k}$ that needs a c-electron to screen a spin impurity obeys:
\begin{equation}
{\tau_k}=\frac{\hbar}{T_k^{(imp)}}<<{\tau_h}=\frac{\hbar}{\epsilon(k)}
\end{equation}
the physics for $T<<T_k^{(imp)}$ may be thought as a 
collection of {\bf independent} ``localization clouds'', inside which a single conduction electron is {\it trapped} by $z>>1$ magnetic 
impurities (cf Fig.5B). $\tau_h$
is the large time required for a Fermi electron to jump on neighboring 
sites. At first level, one can omit the hopping term 
($\tau_h\rightarrow +\infty$) to discuss irrelevant operators in a given
localization cloud. Away from half-filling, only the electrons near the Fermi level are submitted to a magnetic Kondo localization. 

Then, one can make a simple relation of duality between the 
one impurity Kondo model and the one-dimensional Kondo lattice one (cf Fig.5B). In each localization cloud, successive spin flips with the same conduction electron located at $x=x_{o,i}$, are attempted to generate ``similar'' strong 
interactions between spins of the array, as those between c-electrons in the 
one-impurity Kondo model. As in the one-impurity Kondo model, these spin interactions
are inevitably {\it antiferromagnetic}. Inside a given 
cloud $i$ around a localized c-electron at $x=x_{o,i}$, the following ``2-particle''
interactions may be engendered: 
\begin{equation}
\delta{\cal H}_s=-\eta\ f^{\dag}_{\uparrow}(x)f_{\uparrow}(x)f^{\dag}_{\downarrow}(x)f_{\downarrow}(x)
\end{equation}
 Here, it is the single permitted irrelevant operator since the f-electrons 
are 
 frozen on lattice sites $(U_f\rightarrow +\infty)$. Unlike the one-impurity Kondo model, one cannot successfully use the non-Abelian bosonization 
formalism. To explore the nature of the strong-coupling 
regime, we just employ the well-known pseudo-fermion representation of the 
local moments $\vec{\tau}(x)=\frac{1}{2}f^{\dag}(x)\vec{\sigma}f(x)$.
\\
\begin{figure}[ht]
\centerline{\epsfig{file=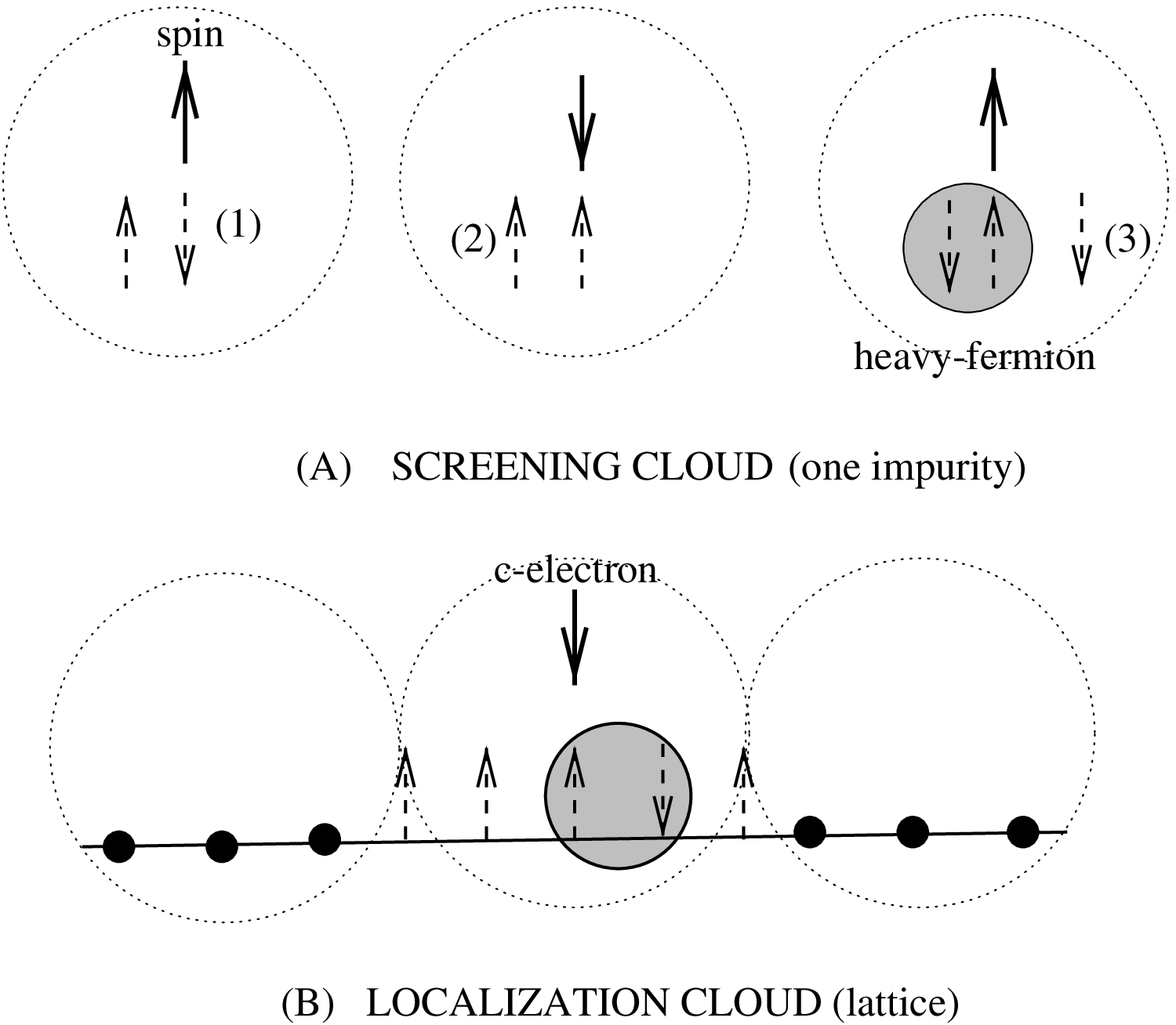,angle=0.0,height=6cm,width=7cm}}
\vskip 1cm
  \caption{{\bf (A)} In the one-impurity Kondo model, successive spin flips
with the localized spin lead to residual strong spin interactions between 
electrons. {\bf (B)} In a given localization 
cloud of the Kondo lattice, successive spin flips with the same c-electron 
should also engender strong interactions between spins of the array.}
 \end{figure}

The following calculations do not consist of a rigorous proof, but rather suggest a
new physical approach of the 1D KLM, away from half-filling. $\eta$ is now expected to play a crucial role. As in the one-impurity Kondo model (Appendix C), we obtain that:
\begin{eqnarray}
\eta&=&\frac{\text{Range scale of irrelevant operators}}{\text
{Fixed length scale}}\\ \nonumber
&=&\frac{\xi_{i}}{L}=\frac{1}{T_k^{(imp)}}\delta(x-x_{o,i})
\end{eqnarray}
To probe irrelevant operators, we have to consider a particular 
localization cloud. Then, since we have the scaling argument $L=1/\left|k-k_F\right
|>>\xi_i$, interactions in a given localization cloud can be considered
as $\delta$-functions\cite{seize}. It confirms that strong quantum fluctuations would be the key to generate ``high-energy'' irrelevant processes at very low-energy. It leads to strong short-range spin-spin correlations:
\begin{eqnarray}
\langle\vec{\tau}(x)\vec{\tau}(0)\rangle&\sim& const.\qquad \text{for}\ x<\xi_i\\ \nonumber
\langle\vec{\tau}(x)\vec{\tau}(0)\rangle&\sim& \exp(-\xi_i/x)\qquad \text{for}\ x>\xi_i\
\end{eqnarray}
This picture allows to give a simple relationship between the Kondo screening
and the magnetic polarization of the spin array. Spatial correlations inside
a localization cloud are assumed to be very strong. Rigorously, one can suppose a small spatial modulation of the spin correlations at short-distances. Finally, this sheme legitimizes to forget the high-energy electrons in the description of the 1D KLM away from half-filling.

In the 1D HKLM $(J_H=t)$, a metallic
phase with a {\it tiny} spin gap has been also found using the non-Abelian
bosonization technique and confirmed by DMRG results\cite{e}:
\begin{equation}
T_{kh}\sim E_c e^{-2\pi(J_H+t)/J_K}<T_k^{(imp)}
\end{equation}
 In that case, all the spins of the array are coupled through a large $J_H$ 
 exchange and the presence of spinons in the Heisenberg chain tend to prevent
 the occurrence of a spin gaped phase. It explains why a finite
 value of $J_H$ tends to decrease the Kondo gap. When $J_H\not =0$, we also remark that the preceding analysis in terms of independent localization clouds fails. This case is very hard to handle because here the Kondo effect is
a $\delta(z)$ function, where $z=x+v_Ft$ is the Fermi-cone component. In the x-space, it is not obligatory a {\it local} coupling and finally it is difficult to give a sense to the very large length scale $\xi_{kh}\sim v_F/T_{kh}$. But anyway, we do not predict any irrelevant operator in $\eta\rightarrow 1/T_{kh}$. Indeed, due to the Heisenberg exchange term, we have the constraint:
\begin{equation}
\tau_H=\frac{\hbar}{J_H}<<\tau_{kh}=\frac{\hbar}{T_{kh}}
\end{equation} 
As soon as a spin is liberated from the Kondo process, it is immediately re-submitted to the Heisenberg exchange term. In that context, the coherence in the spin array is due to the bare exchange Heisenberg term which rather favors massless spin excitations. This allows the Kondo process to remain relevant at 
zero temperature. The hole-doped 1D HKLM is indisputably ruled by a metallic behavior and a Kondo gap\cite{e}. 

Conversely, for $J_H=0$, one can easily predict that the Kondo 
effect cannot survive at the fixed point, due to these strong residual 
interactions between $\tau$-spins $1/T_k^{(imp)}>>T_k^{(imp)}$. The 
low-energy physics will be inevitably governed by a singlet state formed by the $\tau$-spins. We rather suggest the existence of another energy scale in the Kondo lattice, defining the energy  scale at which {\bf all} the $\tau$-spins mutually screen one another, and finally at which the notion of localization cloud \underline{vanishes}.

\subsection{Heavy-fermion behavior and magnetic instability}

The Kondo energy is quantified and finally, the time required 
to engender a coherent state between all spin impurities can 
be estimated as:
\begin{equation}
\tau_{coh}=z\tau_k
\end{equation}
It is approximatively the time that spends a c-electron to interact with  
$z$ magnetic impurities around it. Each visited spin 
is supposed to immediately give rise to a bond singlet state with another
liberated spin in view to remain delocalized at the Fermi level. Then, using the equality
$T_{coh}=\frac{\hbar}{\tau_{coh}}$, the Kondo description should vanish at\cite{f}:
\begin{equation}
T_{coh}=\frac{1}{t}[T_k^{(imp)}]^2=E_c e^{-4\pi t/J_K}
\end{equation}
In that sense, the Kondo effect engenders a coherence between the 
spins of the array; but, it is different from that obtained at 
half-filling. The coherence is now created by successive spin flips with rare
c-electrons and may only occur at very low-energy. Curiously, we can remark
that the two temperatures $T_{kh}$ and $T_{coh}$ are very close ome from another.

In the one-impurity Kondo model, the essence of the Kondo effect is the 
formation of bond states between localized moments and conduction 
electrons. In the Kondo lattice, away from half-filling, the essence of the 
Kondo effect is rather the formation of a ``singlet bond'' state formed between spins of 
the array. At the coherence, localization clouds disappear and singlet  
bonds can be formed between two spins at a distance up to $x=\xi_{coh}=z^2a>>\xi_i$ one from the other. The range of the spin-spin correlations is then considerably enhanced: $\langle\vec{\tau}(x)\vec{\tau}(0)\rangle \sim const.\ \text{for}\ x<\xi_{coh}$. The spin degrees of freedom of the resultant fluid are essentially those of the f-electrons. Unlike the one-impurity Kondo model, we predict the disappearance of elastic scattering at the Fermi level. The f-electrons are all paired together at the Fermi level, where rare c-electrons are also present. As soon as $T\rightarrow T_{coh}$, the composite object $(\vec{\tau}\vec{\sigma}^{\alpha}_{\beta})c_{\beta}$ is then expected to behave as a single object, represented by its contraction as a single fermionic object f at the Fermi level\cite{g}:
\begin{equation}
(\vec{\tau}\vec{\sigma}^{\alpha}_{\beta})c_{\beta}=f_{\alpha}
\end{equation}
Since the f-electrons are paired together, the {\it contraction} of the Kondo 
exchange term gives finally rise to a resonant hybridization between f and c-electrons:
\begin{equation}
T_k^{(imp)}[c_j^{\dag}(\vec{\sigma}.\vec{\tau})c_j\ +(h.c)]\rightarrow 
T_k^{(imp)}[c_j^{\dag}f_j\ +(h.c)]
\end{equation}
Unlike the weak-coupling regime, the Kondo exchange is ruled by $T_k^{(imp)}$
in the spin gaped phase. It may give another meaning to the energy scale $T_{coh}$. It characterizes
the crossover between the Kondo regime and the following valence fluctuation 
regime. Due to the hybridization term $V_{eff}=T_k^{(imp)}$, the delocalized f-electrons are governed by the {\it one-body} Green function:\begin{equation}
\langle f^{\dag}(x,t)f(x,0)\rangle=\sum_n\frac{e^{iw_n t}}{iw_n-iT_{coh}sgn w_n}
\end{equation}
This leads to a free energy variation:
  \begin{equation}
   \delta F(T)\propto\int\frac{d\omega}{\pi}f(\omega)[\tan^{-1}(\frac{T_{coh}}{\omega})]
    \end{equation}
  $f(\omega)$ is the Fermi distribution function, and finally a heavy-fermion
behavior characterized by the huge specific heat $\delta C/T\propto
  1/T_{coh}$ and susceptibility $\delta\chi\propto 1/T_{coh}$. 
  \\
\begin{figure}[ht]
\centerline{\epsfig{file=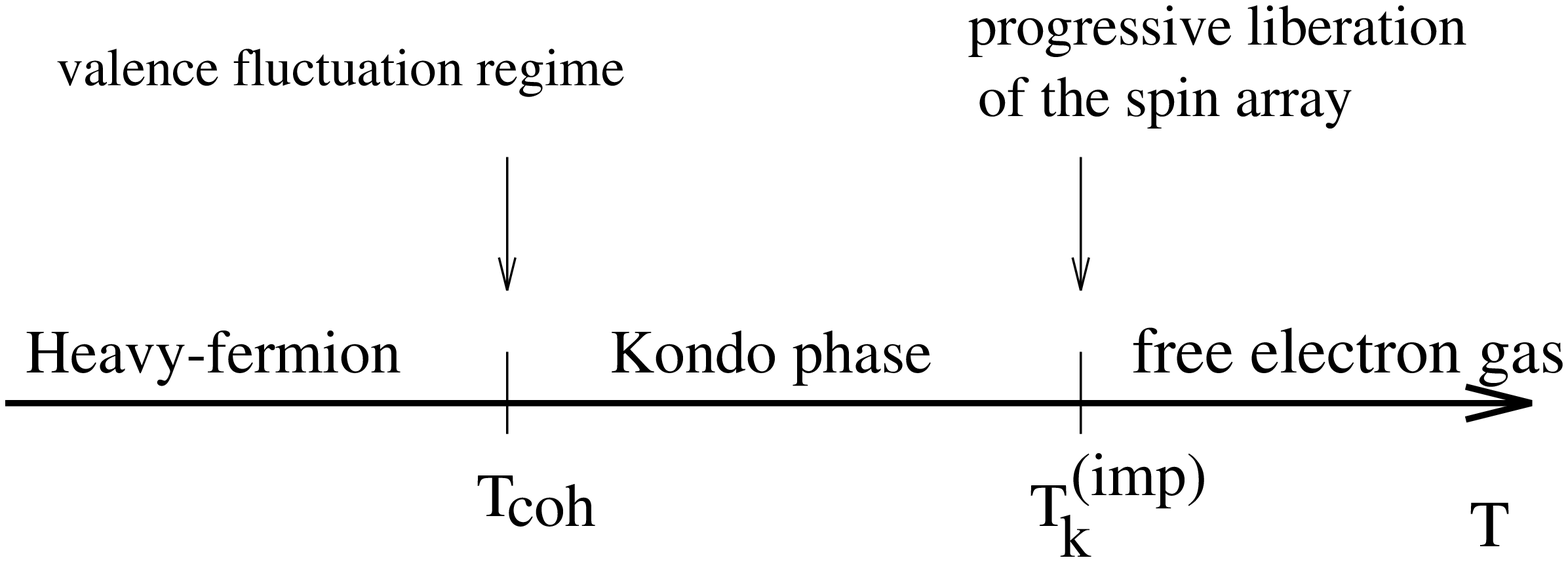,angle=0.0,height=2.5cm,width=7cm}}
\vskip 1cm
 \caption{Cross-over from the Kondo phase to the valence fluctuation regime in
 the Kondo lattice away from half-filling.}
 \end{figure}

The heavy-fermion character is considerably enhanced compared to the 
  one-impurity case. The free energy variation $\delta F(T)$ is in fact the 
  same which rules the fixed point of the Kondo impurity problem in the 
  {\it anisotropic} limit $J_{K\perp}=T_k^{(imp)}<<J_{Kz}$. The energy level
  of the delocalized magnetic impurities is enlarged to $T_{coh}$: this
  problem becomes isomorphic to the well-known resonant level 
  problem. Then, we may predict an 
  universal Wilson ratio:
\begin{equation}
R_W=\frac{C_o}{C_{os}}=\frac{\delta \chi/\chi_o}{\delta C/C_o}=2
\end{equation}
in the Kondo lattice away from half-filling. As in the one-impurity 
problem, it simply measures the ratio of the free electron gas specific heat $C_o=C_{os}+C_{oc}=2C_{os}$ coming from the spin degrees of freedom (Appendix C). 

Here, heavy-fermion quasiparticles are obtained due to the small admixture of
f-electrons at the Fermi level by c-electrons: $\psi_{{\bf k}\sigma}=\cos
{\alpha_{\bf k}}c_{{\bf k}\sigma}+\sin\alpha_{\bf k}
f_{{\bf k}\sigma}$, with the Hamiltonian $H=\sum_{\bf{k}\sigma}\ E_k\ \psi^{\dag}_{{\bf k}\sigma}.\psi_{{\bf k}\sigma}$ and:
\begin{equation}
\tan\alpha(E)=\frac{T_k^{(imp)}}{\left|E\right|}
\end{equation}
We have the two following constraints: 
$\alpha_{\bf k=0}=0$ and $\alpha_{\bf k=k_F}\sim\pi/2$. The wave function can thus be divided up into two parts: a part dominated by
free conduction electrons and a part dominated almost exclusively by 
f-spins. Near $k\rightarrow k_F$, the phase shift is maximum and shows the
prevalence of f-electrons; the scattering is resonant. We finally predict an
enlarged Fermi surface whose enclosed volume counts both the c-electrons and
$\psi$ quasiparticles.

\section{Conclusion}

In this paper, the ground state of the one-dimensional Kondo lattice model
has been completed in the weak-coupling regime, using the non-Abelian bosonization.

At half-filling, we obtain a particular Kondo insulator, namely the Spin Density Glass state. It consists of a $q=\pi$ Spin Density Wave weakly pinned by
a quite antiferromagnetically ordered spin array. 

$\bullet$ The resulting spin system yields a magnetic instability at the wave vector number $q=\pi$. On the other side, this Kondo insulator shows a gap both in charge and spin excitations, which
varies linearly with the Kondo exchange $J_K$ when the electrons are 
non-interacting. On-site repulsion between electrons increases excitation gap 
as $J_K^{2/(3-K_{\rho})}$ where $K_{\rho}=1-U\pi/v_F$ is the charge parameter of the
1D electron gas. At half-filling, backward Kondo scattering processes are 
very coherent and leads to a magnetic like localization of the one-dimensional 
electron gas. The Kondo coupling allows to lift the high degeneracy of the 
impurity lattice by ``ordering'' antiferromagnetically the spin array. We have 
also studied the stability of the SDG state in presence of randomness. The 
Anderson localization by disorder takes place only when the pure 1D electron system yields an enhanced Charge Density Wave. However, it is important to mention that no stable fixed point can be found when conduction electrons are supposed to be non-interacting.

$\bullet$ A large Heisenberg exchange coupling 
between spins of the array destroys the magnetic instability and the 
ground state rather resembles the so-called quantum disordered spin
liquid state of the two-leg spin ladder system. Spin excitations are still 
triplet, but the spin gap or the well-known Haldane gap decreases due to the 
prominence of quantum fluctuations. Such a state is now very sensitive to quenched
disorder.

Away from half-filling, the $2k_F$ oscillation is not commensurate with the lattice and a fascinating metal-insulating transition is previous to take 
place. In the incommensurate limit, the forward Kondo scattering term becomes 
then relevant. As in the one-impurity Kondo model, we observe the existence of 
a Kondo phase ruled by a metallic behavior and the {\it very small} spin gap 
$T_k^{(imp)}\propto\exp(-\pi t/J_K)$, where $t$ is the hopping 
term. Away from half-filling, only conduction electrons near the Fermi level 
are supposed to participate in the Kondo effect and quantum fluctuations 
become crucial. 

$\bullet$ Unlike the strong coupling
regime, the notion of {\it unpaired} localized spin becomes meaningless. The physics in this Kondo phase may be rather thought as a collection of {\it independent} ``localization clouds'' of size $\xi_i=v_F/T_k^{(imp)}$, inside which a single conduction electron is trapped by $t/T_k^{(imp)}$ impurities. In a given localization cloud, successive spin flips with the same
conduction electron generate strong antiferromagnetic interactions between 
spins of the array in $1/T_k^{(imp)}>T_k^{(imp)}$. As a consequence, this 
Kondo phase ruled by singlet bonds between conduction electrons and spins of 
the array cannot survive at the fixed point. 

$\bullet$ We have rather obtained the formation of a singlet bond state
formed between spins of the array at the very low energy scale $T_{coh}\sim
[T_k^{(imp)}]^2/t$. This spin system also yields a quite long-range 
antiferromagnetic polarization. From this point of view, strong coupling in
the Kondo lattice away from half-filling is not so much a binding of conduction electrons to local moments, but rather a {\it liberation} or delocalization of the magnetic moments. Therefore, at the fixed point, the contraction of the Kondo exchange gives finally rise to a resonant Anderson hybridization leading to a normal heavy-fermion state. Finally, the fixed point becomes isomorphic to that of the one-impurity Kondo model in the {\it anisotropic} limit $J_{K\perp}=T_k^{(imp)}<<J_{Kz}$. The analogy with resonant level problems becomes then essential: the spectral width of a liberated impurity is enlarged to $T_{coh}$. 

$\bullet$ Finally, we have formulated scaling arguments to show that this heavy-fermion state is very {\it unstable} by the 
introduction of a Heisenberg exchange between spins of the array. Here, an 
interesting phase with a spin gap but no charge gap subsists at zero 
temperature, as obtained in ref.\cite{e}.
\\

\hskip 2cm {\bf ACKNOWLEDGEMENTS}
\\
\\
The author would like to thank I. Affleck, B. Coqblin and B. Dou\c cot for
helpful criticisms and interesting comments. 
\\

{\bf APPENDIX A: TEMPERATURE DEPENDENCE OF THE CONDUCTIVITY}
\\
\\
For a single impurity $\vec{\tau}$ located at $x=0$, the term $\lambda_3$ should obey\cite{k}:
\begin{equation}
\frac{d\lambda_3}{dln L}=\frac{1}{2}(1-K_{\rho})\lambda_3=\frac{U\lambda_3}{2\pi v_F}
\end{equation}
$\lambda_3$ is here expected to become strong, at the temperature
scale\cite{k}:
\begin{equation}
T_k^{(3o)}\simeq E_c(\frac{J_K}{U})^{2\pi v_F/U}
\end{equation} 
Unlike the Kondo lattice problem, the staggered operator $\lambda_3$ is only relevant for $U>J_K$. In the other cases, we obtain the usual Kondo behavior ruled by the forward scattering term $\lambda_2$. When $U$ is sufficiently 
large, the impurity becomes sensitive to the $2k_F$ polarization in the electron gas, and the magnetic localization length becomes equal to:
\begin{equation}
\xi_{imp}\sim [\frac{1}{\lambda_3}]^{\frac{2}{1-K_{\rho}}}>>\xi_c
\end{equation}
A single magnetic impurity introduced in a TL liquid, do not easily manage 
to trap a $q=2k_F$ SDW. The conductance of the one-dimensional electron gas, in the presence of this single impurity, is already
disturbed for $L<\xi_{imp}$:
\begin{equation}
\delta G_{imp}(T)=G_{imp}(T)-G_o\sim\lambda_3^2(T)\propto T^{K_{\rho}-1}
\end{equation}
The resistance follows ${\cal R}(T)=1/G_{imp}(T)=1/G_o+T^{K_{\rho}-1}$. Suppose now that, in the Kondo lattice, the magnetic impurities were \underline{totally} disconnected in the delocalized phase. It would give the total resistance ${\cal R}_{tot}(T)\sim LT^{K_{\rho}-1}$, and
a conductivity:
\begin{equation}
 \sigma_d(T)=L{\cal R}_{tot}^{-1}(T)=[\delta G_{imp}(T)]^{-1}
\end{equation}
It is not equal to the conductivity $\sigma(T)$ obtained in the Kondo lattice by a direct calculation. It confirms that in the weak-coupling regime the impurities are submitted to short-range polarization effects due to the renormalization of $\lambda_3$, which enhance the DC conductivity. When $U\rightarrow 0$, the conclusion is still more obvious since the $q=0$ Kondo process generates
logarithmic corrections at high energy. 
\\

{\bf APPENDIX B: EFFECTIVE ACTION OF THE HEISENBERG-KONDO CHAIN}
\\
\\
 To study in more details the spin sector, we replace the charge operator by its non-universal value $K=\lambda_3 \langle\cos(\sqrt{2\pi}\Phi_c^c)\rangle$. K varies linearly with $T_k^{(3)}$. To make the calculations simple as 
 possible, we consider the particular constraint 
$J_H=t$. We also use the representation of 
 $\vec{J}_{c,(L,R)}$ in terms of the matrix g:
 \begin{equation}
   \label{deux} 
   \vec{J}_{c,L}=-\frac{i}{2\pi} tr[\partial_- gg^{\dag}.\vec {\sigma}]\qquad \vec{J}_{c,R}=\frac{i}{2\pi} tr[g^{\dag}\partial_+g.\vec {\sigma}]
   \end{equation}
Then, the spin action comes out as:
\begin{eqnarray}
  S&=&W(f)+W(g)\\ \nonumber
&-& \frac{K}{2}\int d{\bf x}\ tr(g^{\dag}.\vec{{\sigma}}) tr(f.\vec{{\sigma}})+(h.c)
  \end{eqnarray}
${\bf x}=(x,v_Ft)$, and $W(f)$ is the following Wess-Zumino-Witten action: 
\begin{equation}
 W(f)=\frac{1}{4 u^2}\int d {\bf x}\ tr\partial_{\mu}f \partial^{\mu}f^{\dag}-\Gamma(f)
   \end{equation}
where $u^2\propto v_F^{-1}$. $\Gamma(f)$ is the topological functional:
 \begin{equation}
\Gamma(f) =\frac{k}{\pi}\int_0^{\infty} d\xi\int d{\bf x}\ \epsilon^{\mu\nu\lambda} tr f^{\dag}\partial_{\mu}f f^{\dag}\partial_{\nu}ff^{\dag}\partial_{\lambda}f
 \end{equation}  
 It is convenient to consider the action rather than the Hamiltonian
 because topological effects are expected to be prevalent in
 such dimensional spin systems. We remind that $k=1$ for a Heisenberg chain as well as a one-dimensional
electron gas; $k$ is not submitted to renormalization, and more especially
it is defined ``modulo 2''. It gives \underline{power laws} for the
spin-spin correlations of both the Heisenberg chain, and the one-dimensional 
electron gas. 

We notice that $S_{int}=-H_{int}$. The form of $S_{int}$ indicates that all
spin modes are gaped. Now, we substitute $(f=g\alpha^{\dag})$ in the above action in 
view to separate singlet and triplet excitations in the spectrum. We use
the identity:
\begin{equation}
tr(f^{\dag}.\vec{\sigma})tr(g.\vec{\sigma})=\frac{1}{2}tr(f^{\dag}.g)-\frac{1}{4}trf^{\dag}.trg
\end{equation}
and the approximation:
\begin{equation}
trf^{\dag}.trg\sim \sqrt{T_k^{(3)}}.trg-tr\alpha
\end{equation}
By construction, the terms in $trg$ and $tr\alpha$ have an opposite 
sign. Then, the interacting part of the action can be summed up into:
\begin{equation}
S_{int}=-\frac{3}{4}T_k^{(3)}tr\alpha+\frac{1}{2}{[T_k^{(3)}]}^{3/2}.trg
\end{equation}
With the Abelian bosonization language, we obtain that:
\begin{equation}
tr\alpha=tr(f^{\dag}g)\sim 2\cos(\sqrt{4\pi}\phi_-)+ 2\cos(\sqrt{4\pi}\tilde{\phi}_-)
\end{equation}
It confirms that $tr\alpha$ has the scaling dimension 1 (and not
1/2) and then $\langle tr\alpha\rangle\sim T_k^{(3)}$. Using the Majorana fermionic representation $(\rho,\vec{\xi})$ introduced
 by the authors of ref.\cite{b} to distinguish triplet $\vec{\xi}$-excitations and singlet $\rho$-ones, we find
that the operator $-\frac{3}{4}T_k^{(3)}tr\alpha$ simply gives the
(expected) mass $M=-3T_k^{(3)}$ for the singlet branch $\vec{\rho}$. We conclude that, first, the $\alpha$ field takes into account high-energy
spin excitations only. Second, with the used $(\alpha,g)$ non-Abelian 
representation, we have completely separated the singlet and triplet 
excitations. 

At long distances ($L>>[T_k^{(3)}]^{-1}$), the matrices $f$ and $g$ are
traceless: the system behaves as a system of two coupled identical 
spin chains. The asymptotic behavior should be governed by an action of type
2W(g), defined by $\tilde{u}=u/\sqrt{2}$, and the topological number $k=0\ [mod\  2]$. To prove that explicitly, it is enough to remark that:
 \begin{eqnarray}
  W(f)=W(g\alpha^{\dag})&=& W(g)+W(\alpha)\\ \nonumber
&+&\frac{1}{8\pi} tr \int d{\bf x}\ g^{\dag}\partial_+g\alpha^{\dag}\partial_-\alpha
  \end{eqnarray}
 Although the action $W(\alpha)$ is ruled by $k=0$, we can check that its
kinetic term has no influence on the low-energy spin 
excitations: $u^{-1}<<\tilde{u}^{-1}$. A coupling like 
$(\partial_-\alpha.g^{\dag})$ cannot be no longer relevant at long distances. 
\\
The low-energy spin sector is finally well-modeled by an action of type:
\\
\begin{equation}
\label{un}
S_{eff}=\frac{2}{4 u^2}\int d {\bf x}\ tr\partial_{\mu}g \partial^{\mu}
g^{\dag}-\gamma{[T_k^{(3)}]}^{3/2}trg
\end{equation}
\\
$\gamma=1/2$ with the used conventions. Due to the massive term $trg$, the invariance
per translation is now broken. This effective action remains available in the
case of the spin S=1 chain; in that context, $T_k^{(3)}\sim t=J_H$.
\\

{\bf APPENDIX C: SCREENING CLOUD IN THE IMPURITY KONDO PROBLEM}
\\
\\
In the one-impurity Kondo model, the electrons participating in the formation 
of a singlet located at the impurity site may lie within a distance 
$\xi_i=v_F/T_k^{(imp)}$. This is a length that is far greater than the lattice spacing, and finally it forms a real ``screening cloud'' around
the impurity site. The low-behavior seems trivial; the impurity has disappeared
from the low-energy physics. But, certains interactions between 
electrons are generated in the processes of eliminating the impurity spin. 
The fixed point analysis can be made using a one-dimensional mapping of the Kondo
Hamiltonian, formulated with left movers only. Then, it has been shown that much of the interesting behavior comes from the two following {\it irrelevant} operators\cite{quinze,seize}:
\begin{equation}
\delta{\cal H}=\lambda_1\ ic^{\dag}_{L\alpha}\partial_x c_{L\alpha}\delta(x)-\lambda_2\ c^{\dag}_{L\uparrow}c_{L\uparrow}c^{\dag}_{L\downarrow}c_{L\downarrow}\delta(x)
\end{equation}
An irrelevant operator is defined by a scaling dimension d>1. In 1D, the scaling dimension of a fermion is $[c]=1/2$. $\lambda_1$ and $\lambda_2$ are then
irrelevant operators and $[\lambda_i]=-1$, meaning that they vary as the 
inverse of an energy. Since we investigate the physics at a distance $L\rightarrow +\infty$, interactions in the screening cloud can be modeled by 
$\delta$ functions. Nozi\` eres\cite{quinze} argued that they have an universal ratio $R_W$ (``the Wilson number''); there is a single unknown parameter. To 
keep the invariance per rotation, Affleck\cite{seize}
finds that $\lambda_1/\lambda_2=2=R_W$ and finally, $\delta{\cal H}$ can be
summed up into the spin current term:
\begin{equation}
\delta{\cal H}=\delta{\cal H}_s=\lambda_1:\vec{J}_{c,L}({x})\vec{J}_{c,L}({x}):\delta(x)
\end{equation}
Since this analysis is only correct for $T<T_k^{(imp)}$, we deduce that
$\lambda_1\propto 1/T_k^{(imp)}$. At
the fixed point, the spin-charge phenomenon survives but strong spin-spin
interactions in the screening cloud rescales:
\begin{equation}
\vec{J}_{L}(x)\rightarrow\vec{J}_{L}(x)\sqrt{1+\frac{3\lambda_1}{2\pi v_F L}}
\end{equation}
where $\delta(x)\sim 1/L$. Finally, the termodynamics is really
affected by the strong electronic correlations in the screening cloud:
\begin{eqnarray}
\chi(T,\lambda_1)&=&\frac{1}{3T}\langle[\int dx\ \vec{J}_{L}(x)]^2\rangle_{\lambda_1}\\ \nonumber
                 &=&\frac{[1+\frac{3\lambda_1}{2\pi v_F L}]}{3T}\langle[\int\ dx\ \vec{J}_{L}(x)]^2\rangle_{\lambda_1=0}\\ \nonumber
                 &\simeq&[1+\frac{3\lambda_1}{2\pi v_F L}]\chi_o=\chi_o+\delta\chi
\end{eqnarray}
and,
\begin{eqnarray}
 C_s(T,\lambda_1)&=&\frac{\partial}{\partial T}\langle \int dx\ {\cal H}_s\rangle_{\lambda_1}\\ \nonumber
                &=&(1+\frac{3\lambda_1}{2\pi v_F L})\frac{\partial}{\partial T}\langle \int dx\ {\cal H}_s\rangle_{\lambda_1=0}\\ \nonumber
                &=&(1+\frac{3\lambda_1}{2\pi v_F L})C_{os}=C_{os}+\delta C\\ \nonumber
\end{eqnarray}
where $\chi_o=L/2\pi v_F$ and $C_{os}=\pi L/3 v_F$ are the susceptibility and the specific heat coming
from the spin degrees of freedom in the free system; they vary as 
$L/v_F$. We obtain a heavy-fermion behavior ruled by $\delta C/T\propto 1/T_k^{(imp)}$
and $\delta\chi\propto 1/T_k^{(imp)}$. The Wilson number
 can be finally reexpressed as: 
\begin{equation}
R_W=\frac{C_o}{C_{os}}=\frac{\delta \chi/\chi_o}{\delta C/C_o}=2
\end{equation}
It simply measures the ratio of the total specific heat 
$C_o=C_{os}+C_{oc}=2C_{os}$ coming from the spin degrees of 
freedom. 

Finally, due to the importance of quantum spin fluctuations, a single located 
impurity manages to capture all c-electrons near the Fermi level. It 
modifies drastically the low-energy physics of the 1D electron
gas: bare free electrons turn into free heavy-fermions. Finally, we want to
insist on the fact that the fixed point properties are simply ruled by the
dimensionless ``magic'' number:
\begin{equation}
\eta=\frac{\xi_{i}}{L}=\frac{\text{Range scale of irrelevant operators}}{\text{Fixed length scale}}
\end{equation}


\begin{references}
\bibitem{un} For a complete review, see H. Tsunetsugu, M. Sigrist and K. Ueda, Rev. Mod. Phys. vol {\bf 69}, p809-864 (1997).
\bibitem{deux} M.F. Hundley, P.F. Canfield, J.D. Thompson, Z. Fisk and
J.M. Lawrence, Phys. Rev. B {\bf 42}, 6842 (1990).
\bibitem{trois}  B. Coqblin, J. Arispe, J.R. Iglesias, C. Lacroix and K. Le 
Hur, J. Phys. Soc. Jpn {\bf 65}, Suppl. B 64 (1996). A.C Hewson, {\it The Kondo problem to heavy Fermions} (Cambridge University press, Cambridge, 1993). For a more general review on heavy-fermion physics, see N. Grewe and F. Steglich, {\it Handbook on the Physics and
Chemistry of Rare Earths}, eds. K.A. Gschneider  and L. Eyring, {\bf 14}, 343 (1974).
\bibitem{quatre} R. Jullien and P. Pfeuty, J. Phys. F {\bf 11}, 353 (1981).
\bibitem{cinq} L.G. Caron and C. Bourbonnais, Europhys. Lett. {\bf 11}, 473 (1990).
\bibitem{six} H. Tsunetsugu, Y. Hatsugai, K. Ueda and M. Sigrist, Phys. Rev. B {\bf 46}, 3175 (1992).
\bibitem{sept} R.M. Fye and D.J. Scalapino, Phys. Rev. Lett. {\bf 65}, 3177 (1990).
\bibitem{huit} M. Troyer and D. W\" urtz, Phys. Rev. B {\bf 47}, 2886 (1993).
\bibitem{neuf} C.C. Yu and S.R. White, Phys. Rev. Lett. {\bf 71}, 3866 (1993).
\bibitem{neuf1} A.M. Tsvelik,  Phys. Rev. Lett. {\bf 72}, 1048 (1994).
\bibitem{dix} {\bf Abelian bosonization}: See, e.g, V.J. Emery, in {\it Highly Conducting One-Dimensional Solids}, edited by J.T. Devreese, R.P. Evrard and V.E. Van.Doren, (Plenum,New York,1979), p.327; J. Solyom, Adv.Phys. {\bf 28}, 201 (1979); S. Sachdev and R. Shankar, Phys. Phys. Rev. B {\bf 38}, 826 (1988). 
\bibitem{dix1}  Concerning the 1D KLM, we suggest: S.P. Strong and A.J. Millis, Phys. Rev. Lett. {\bf 69}, 2419 (1992); S. Fujimoto and N. Kawakami, J. Phys. Soc. Jpn  {\bf 63}, 4322 (1994).
\bibitem{dix2} O. Zachar, S.A. Kivelson and V.J. Emery, Phys. Rev. Lett. {\bf
77}, 1342 (1996).
\bibitem{onze} {\bf Non-Abelian bosonization}: See I. Affleck in {\it Fields, Strings and Critical phenomena} edited by E. Brezin and J. Zinn-Justin (North-Holland, Amsterdam, 1990).
\bibitem{onze1} Concerning the 1D KLM, we suggest: S.R. White and I. 
Affleck, Phys. Rev. B {\bf 54}, 9862 (1996); K. Le Hur, Phys. Rev. B {\bf 56}, 14056 (1997).
\bibitem{d} Z. Wang, X.P. Li and D.H. Lee,  Phys. Phys. Rev. B {\bf 47}, 11935 
(1993).
\bibitem{douze1} See R. Shankar, Rev. Mod. Phys. {\bf 66}, 129 (1994).
\bibitem{douze2} S. Tomonaga, Prog. Theor. Phys. (Kyoto) {\bf 5}, 544 (1950); J.M. Luttinger, J. Math. Phys. N.Y. {\bf 4}, 1154 (1963); F.D.M. 
Haldane, J. Phys. C{\bf 14}, 2585 (1981); H.J. Schulz in
{\it Mesoscopic Quantum Physics, Les Houches, Session LXI, 1994}, edited by E. Akkermans, G. Montambaux, J.L. Pichard
                  and J. Zinn-Justin (Elsevier, Amsterdam, 1995), p. 533.
\bibitem{vingt} S. Moukouri and L.G. Caron, Phys. Rev. B {\bf 54}, 12212 (1996).
\bibitem{trente} N. Shibata, A. Tsevlik and K. Ueda, Phys. Rev. B {\bf 56}, 330 (1997).
\bibitem{douze} G. Honner and M. Gul\' acsi,  Phys. Rev. Lett. {\bf 78}, 2180 (1997).
 \bibitem{zero} N. Shibata, T. Nishino, K. Ueda and C. Ishii, cond-mat/9608117.
 \bibitem{treize} T. Giamarchi and H.J. Schulz, Phys. Rev. B {\bf 37}, 325 (1988).
\bibitem{quatorze} S.F. Edwards and P.W. Anderson, J. Phys. F {\bf 5}, 965 (1975).
\bibitem{quinze}  K.G. Wilson, Rev. Mod. Phys. {\bf 47}, 773 (1975); Ph. Nozi\` eres, J. Low Temp. Phys. {\bf 17}, 31 (1974); Ph. Nozi\` eres, {\it Proceedings of the 14th international Conference on Low Temperature physics}, edited by M. Krusius and M. Vuorio (North Holland, Amsterdam,1975), Vol.5.
\bibitem{seize} {\bf Applications of Field Theories to the Kondo
effect}, See I. Affleck cond-mat/9512099.
\bibitem{a} E. Witten, Commun. Math. Phys. {\bf 92} (1984) 455-47.
\bibitem{b} D.G. Shelton, A.A. Nersesyan and A.M. Tsvelik, Phys. Rev. B {\bf 53}, 8521 (1996).
\bibitem{c} Ph. Nozi\` eres, Ann. Phys. Fr {\bf 10}, 19 (1985).
\bibitem{e} A.E. Sikkema, I. Affleck and S.R. White, Phys. Rev. Lett. 
{\bf 79}, 929 (1997).
\bibitem{f} Ph. Nozi\` eres (unpublished) has suggested the relevance of such
an energy sale $T_{coh}\propto [T_k^{(imp)}]^2/t$ in the Kondo lattice, away
from half-filling.
\bibitem{g} P. Coleman, A.M. Tsvelik, N. Andrei and H.Y. Kee, 
cond-mat/9707002.
\bibitem{k} K. Le Hur, cond-mat/9802298.
\end{references}
\end{document}